\documentclass[preprint,authoryear,3p]{elsarticle}
\usepackage{amsthm}
\usepackage{amsmath, mathtools, dsfont}
\usepackage{ bbold }
\usepackage{adjustbox}
\usepackage{physics}
\usepackage{algorithm}
\usepackage{bm}
\usepackage{multirow}
\usepackage{longtable}
\usepackage{ulem}
\usepackage{algpseudocode}
\usepackage{graphicx}
\usepackage[english]{babel}
\usepackage[utf8]{inputenc}
\usepackage{amssymb}
\usepackage{float}
\usepackage{multicol}
\usepackage{booktabs}
\usepackage{mathrsfs}
\usepackage{bbm}
\usepackage{rotating}
\usepackage{makecell}
\usepackage{lscape}
\usepackage{subfigure}
\usepackage[colorinlistoftodos]{todonotes}
\usepackage{lipsum}
\usepackage{comment}

\numberwithin{equation}{section}
\usepackage[figurename=Fig.]{caption}
\theoremstyle{plain}

\makeatletter
\def\ps@pprintTitle{%
  \let\@oddhead\@empty
  \let\@evenhead\@empty
  \let\@oddfoot\@empty
  \let\@evenfoot\@oddfoot
}
\makeatother

\usepackage{hyperref}
\hypersetup{
  colorlinks   = true,    % Colours links instead of ugly boxes
  urlcolor     = jaam,    % Colour for external hyperlinks
  linkcolor    = firebrick,    % Colour of internal links
  citecolor    = blue      % Colour of citations
}

\usepackage[nameinlink,capitalise]{cleveref}
\begin{document}
\begin{frontmatter}
\title{Model-free time-aggregated predictions for econometric datasets\tnoteref{t1}}
\tnotetext[t1]{Declarations of interest: none.}
%This research did not receive any specific grant from funding agencies in the public, commercial, or not-for-profit sectors.
%This work was supported by 
\author[1]{Kejin Wu} \ead{kwu@ucsd.edu} \author[2]{Sayar Karmakar\corref{cor1}} \ead{sayarkarmakar@ufl.edu} \cortext[cor1]{Corresponding author} \address[1]{University of California San Diego, La Jolla, CA, U.S. 92093} \address[2]{University of Florida, Gainesville, FL, U.S. 32611}

\begin{abstract}
This article explores the existing normalizing and variance-stabilizing (NoVaS) method on predicting squared log-returns of financial data. First, we explore the robustness of the existing NoVaS method for long-term time-aggregated predictions. Then we develop a more parsimonious variant of the existing method. With systematic justification and extensive data analysis, our new method shows better performance than current NoVaS and standard GARCH(1,1) methods on both short- and long-term time-aggregated predictions.
\end{abstract}


\end{frontmatter}

\newpage
\section{Introduction}\label{sec:intro}
\noindent Accurate and robust volatility forecasting is a central focus in financial econometrics. This type of forecasting is crucial for practitioners and traders to make decisions in risk management, asset allocation, pricing of derivative instrument and strategic decisions of fiscal policies, etc. Standard methods to do volatility forecasting are typically built upon applying GARCH-type models to predict squared financial log-returns. With the Model-free Prediction Principle, first proposed by \citet{politis2003normalizing}, a model-free volatility prediction method--NoVaS has been proposed recently for efficient forecasting without the assumption of normality. Some previous studies have shown that the NoVaS method possesses better pseudo-out of sample (POOS) forecasting performance than GARCH-type models on forecasting squared log-returns \citep{gulay2018comparison,chen2019optimal}. 

However, to the best of our knowledge, such methods were not evaluated for time-aggregated prediction. Time-aggregated prediction here stands for the prediction of $Y_{n+1}+\cdots +Y_{n+h}$ after observing $\{Y_t\}_{t=1}^{n}$. Such predictions remain crucial for strategic decisions implemented by commmodity or service providers, (\cite{chudy2020long,karmakar2020long}), trust funds, pension management, insurance companies, portfolio management of specific derivatives (\cite{kitsul2013economics}) and assets (\cite{bansal2016risks}). In this paper we will focus on forecasting squared log-returns of different econometric datasets. A time-aggregated forecasting is able to provide somewhat confidence at understanding the general trend in near future, maybe the entire next week or months ahead and definitely stands more meaningful than just understanding what might happen for any single-step ahead (Predicting $Y_{n+h}$ for one value of $h$) in time horizon. In fact, quality of forecasts for econometric data has been evaluated through such time-aggregated metrics in \citep{starica2003garch,fryzlewicz2008normalized}. Apart from exploring the capabilities of existing NoVaS method towards time-aggregated forecasting, we also attempt to improve the existing one further by proposing a more parsimonious model. We substantiate our proposal by extensive simulations and data analysis.

\section{Method}\label{sec:method}
\subsection{The existing NoVaS method}
\noindent NoVaS method is a Model-free prediction principle. The main idea lies in applying an invertible transformation $H$ which can map the non-$i.i.d$. vector $\{Y_i\}_{i=1}^t$ to a vector $\{\epsilon_i\}_{i=1}^t$ that has $i.i.d.$ components. This leads to the prediction of $Y_{t+1}$ by inversely transforming the prediction of $\epsilon_{t+1}$  \citep{politis2015modelfreepredictionprinciple}. The starting point to build the transformation of the existing NoVaS method is the ARCH model \citep{engle1982autoregressive}. Then, \citet{politis2003normalizing} made some adjustments to determine the final form of $H$ as:
\begin{equation}\label{2e2}
      W_{t}=\frac{Y_t}{\sqrt{\alpha s_{t-1}^2+a_0Y_t^2+\sum_{i=1}^pa_iY_{t-i}^2}} ~~\text{for}~ t=p+1,\cdots,n 
\end{equation}

In \cref{2e2}, $\{Y_t\}_{t=1}^n$ is the log-returns vector in this article; $\{W_{t}\}_{t=p+1}^n$ is the transformed vector which we want to make it be $i.i.d.$; $\alpha$ is a fixed-scale invariant constant; $s_{t-1}^2$ is calculated by $(t-1)^{-1}\sum_{i=1}^{t-1}(Y_i-\mu)^2$, with $\mu$ being the mean of $\{Y_i\}_{i=1}^{t-1}$. For reaching a qualified transformation function, \cref{2e3} is required to stabilize the variance.
\begin{equation}\label{2e3}
    \alpha\in (0,1), a_i\geq0~\text{for all}~i\geq0, \alpha + \sum_{i=0}^pa_i=1 
\end{equation} 
Then, $\alpha$ and $a_0,\cdots,a_p$ are finally determined by minimizing $|Kurtosis(W_t)-3|$\footnote{The transformed $\{W_t\}$ is usually uncorrelated, see \citep{politis2015modelfreepredictionprinciple} for additional processes for correlated $\{W_t\}$.}. This method is model-free in the sense, we don't assume any particular distribution for the innovation $\{W_t\}$ except for matching it's kurtosis to 3. Once $H$ is found, $H^{-1}$ can be gotten immediately. For example, $H^{-1}$ corresponding with \cref{2e2} is:
\begin{equation}\label{3.3e1} 
     Y_t=\sqrt{\frac{W_{t}^2}{1-a_0W_{t}^2}(\alpha s_{t-1}^2+\sum_{i=1}^pa_iY_{t-i}^2)}~~\text{for}~ t=p+1,\cdots,n 
\end{equation}
To get the prediction of $Y_{n+1}^2$, \citet{politis2015modelfreepredictionprinciple} defined two types of optimal predictors under $L_1$ (Mean
Absolute Deviation) and $L_2$ (Mean
Squared Error) criterions after observing historical information set $\mathscr{F}_{n} = \{Y_t,1\leq t \leq n\}$:
\begin{equation}\label{Eq.2.4}
\begin{split}
     L_1\text{-optimal predictor of}~&Y_{n+1}^2:\\ &\text{Median}\left\{Y_{n+1,m}^2:m=1,\cdots,M\big\rvert \mathscr{F}_{n}\right\} \\ 
     &= \text{Median}\left\{\frac{W_{n+1,m}^2}{1-a_0W_{n+1,m}^2}(\alpha s_{n}^2+\sum_{i=1}^pa_iY_{n+1-i}^2): m=1,\cdots,M \bigg\rvert \mathscr{F}_{n}\right\} \\
    &=(\alpha s_{n}^2+\sum_{i=1}^pa_iY_{n+1-i}^2)\text{Median}\left\{\frac{W_{n+1,m}^2}{1-a_0W_{n+1,m}^2}:m=1,\cdots,M\right\} \\
    L_2\text{-optimal predictor of}~&Y_{n+1}^2:\\
    &\text{Mean}\left\{Y_{n+1,m}^2:m=1,\cdots,M \big\rvert \mathscr{F}_{n}\right\} \\
    &= \text{Mean}\left\{\frac{W_{n+1,m}^2}{1-a_0W_{n+1,m}^2}(\alpha s_{n}^2+\sum_{i=1}^pa_iY_{n+1-i}^2):m=1,\cdots,M\bigg\rvert \mathscr{F}_{n}\right\} \\
    &=(\alpha s_{n}^2+\sum_{i=1}^pa_iY_{n+1-i}^2)\text{Mean}\left\{\frac{W_{n+1,m}^2}{1-a_0W_{n+1,m}^2}:m=1,\cdots,M\right\} 
    \end{split}
\end{equation}
where, $\{W_{n+1,m}\}_{m=1}^{M}$ are generated $M$ times from its empirical distribution or a normal distribution\footnote{More details about this procedure and multi-step prediction are presented in \cref{susec:newmethods}. Normal distribution here is an asymptotic limit of the empirical distribution of $\{W_{n+1}\}$.}. $\{Y^{2}_{n+1,m}\}_{m=1}^{M}$ are given by plugging $\{W_{n+1,m}\}_{m=1}^{M}$ into \cref{3.3e1} and setting $t$ as $n+1$. During the optimization process, different forms of unknown parameters in \cref{2e3} are applied so that various NoVaS methods were established. \citet{chen2018prediction} pointed out that the Generalized Exponential NoVaS (GE-NoVaS) method with exponentially decayed unknown parameters presented in \cref{2e4} is superior than other NoVaS-type methods. 
\begin{equation}\label{2e4}
    \alpha \neq 0, a_i = c'e^{-ci}~\text{for all}~0\leq i\leq p,~c' = \frac{1-\alpha}{\sum_{i=0}^pe^{-ci}}  
\end{equation} 

\subsection{A new method with less parameters}\label{susec:newmethods}
\noindent 
However, during our investigation, we found that the GE-NoVaS method returns extremely large predictions under $L_2$ criterion sometimes. A \textbf{removing-$a_0$} idea is proposed to avoid such issue in this article. $H$ and $H^{-1}$ of the GE-NoVaS-without-$a_0$ method can be rewritten as below:
\begin{equation}
     W_{t}=\frac{Y_t}{\sqrt{\alpha s_{t-1}^2+\sum_{i=1}^pa_iY_{t-i}^2}}~;~Y_t=\sqrt{W_{t}^2(\alpha s_{t-1}^2+\sum_{i=1}^pa_iY_{t-i}^2)}~;~\text{for}~ t=p+1,\cdots,n \label{2e6} 
\end{equation}

We should notice that even without $a_0$ term, the causal prediction rule is still satisfied. It is easy to get the analytical form of the first-step ahead $Y_{n+1}$, which can be expressed as below:
\begin{equation}
    Y_{n+1}=\sqrt{W_{n+1}^2(\alpha s_{n}^2+\sum_{i=1}^pa_iY_{n+1-i}^2)} \label{3e10}
\end{equation}
More specifically, when the first-step GE-NoVaS-without-$a_0$ prediction is performed, $\{W^*_{n+1}\}$ are generated $M$ (i.e., 5000 in this article) times from a standard normal distribution by Monte Carlo method or bootstrapping from its empirical distribution $\hat{F}_w$\footnote{$\hat{F}_w$ is calculated from \cref{2e2}, i.e., the empirical distribution of transformed series $\{W_t\}_{p+1}^{n}$ corresponding with $\{Y_t\}_{t=1}^{n}$.}. Then, plugging these $\{W^{*}_{n+1,m}\}_{m = 1}^{M}$ into \cref{3e10}, $M$ pseudo predictions $\{\hat{Y}^{*}_{n+1,m}\}_{m=1}^{M}$ are obtained. According to the strategy implied by \cref{Eq.2.4}, we choose $L_1$ and $L_2$ risk optimal predictors $\hat{Y}_{n+1}^2$ as the sample median and mean of $\{\hat{Y}^{*}_{n+1,1},\cdots,\hat{Y}^{*}_{n+1,M}\}$, respectively. We can even predict the general form of $Y_{n+h}$, such as $g(Y_{n+h})$ by adopting the sample mean or median of $\{g(\hat{Y}^{*}_{n+1,1}),\cdots,g(\hat{Y}^{*}_{n+1,M})\}$. Similarly, the two-steps ahead $Y_{n+2}$ can be expressed as:
\begin{equation}
    Y_{n+2}= \sqrt{W_{n+2}^2(\alpha s_{n+1}^2+ a_1Y_{n+1}^2 + \sum_{i=2}^pa_iY_{n+2-i}^2)}
\end{equation}
When the prediction of $Y_{n+2}$ is required, $M$ pairs of $\{W^*_{n+1}, W^*_{n+2}\}$ are still generated by bootstrapping or Monte Carlo method from empirically or standard normal distributions, respectively. $Y_{n+1}^2$ is replaced by the predicted value $\hat{Y}_{n+1}^2$ which is derived from running the first-step GE-NoVaS-without-$a_0$ prediction with simulated $\{W^{*}_{n+1,m}\}_{m=1}^{M}$ under $L_1$ or $L_2$ criterion. Subsequently, we choose $L_1$ and $L_2$ risk optimal predictors of $Y_{n+2}$ as the sample median and mean of $\{\hat{Y}^{*}_{n+2,1},\cdots,\hat{Y}^{*}_{n+2,M}\}$

Finally, iterating the process described above, we can accomplish multi-step ahead NoVaS predictions. $Y_{n+h},h\geq 3$ can be expressed as:
\begin{equation}
    Y_{n+h}= \sqrt{W_{n+h}^2(\alpha s_{n+h-1}^2+\sum_{i=1}^pa_iY_{n+h-i}^2)}
\end{equation}
To get the prediction of $Y_{n+h}$, we generate $M$ number of $\{W^*_{n+1},\cdots,W^*_{n+h}\}$ and plug $\{Y_{n+k}\}_{k=1}^{h-1}$ with NoVaS predicted values $\{\hat{Y}_{n+k}\}_{k=1}^{h-1}$ which are computed iteratively. $L_1$ and $L_2$ risk optimal predictors of $Y_{n+h}$ are computed by the sample median and mean of $\{\hat{Y}^{*}_{n+h,1},\cdots,\hat{Y}^{*}_{n+h,M}\}$. In short, we can summarize that $Y_{n+h}$ is determined by:
\begin{equation}
    Y_{n+h} = f_{\text{GE-NoVaS-without}-a_0}(W_{n+1},\cdots,W_{n+h},\mathscr{F}_{n}) \label{3e12}
\end{equation}
Since $\mathscr{F}_{n}$ is the observed information set, we can simplify the expression of $Y_{n+h}$ as:
\begin{equation}
    Y_{n+h} = f_{\text{GE-NoVaS-without}-a_0}(W_{n+1},\cdots,W_{n+h}) \label{3e13}
\end{equation}
For applying the GE-NoVaS method, we can still build the relationship between $Y_{n+h}$ and $\{W_{n+1},\cdots,W_{n+h}\}$ as:
\begin{equation}
    Y_{n+h} = f_{\text{GE-NoVaS}}(W_{n+1},\cdots,W_{n+h}) \label{3e14}
\end{equation}

We should notice that simulated $\{W^{*}_{n+1,m},\cdots,W^{*}_{n+h,m}\}_{m=1}^{M}$ for obtaining GE-NoVaS method prediction of $Y_{n+h}$ should be generated by bootstrapping or Monte Carlo method from empirically or a trimmed standard normal distribution\footnote{The reason of using the trimmed distribution is $|W_t|\leq1/\sqrt{a_0}$ from \cref{2e2}.}. Here, we summarize the Algorithm \ref{algori1} to perform $h$-step ahead time-aggregated prediction using GE-NoVaS-without-$a_0$ method. The algorithm of GE-NoVaS can be written out similarly.

\begin{algorithm}[htbp]
\caption{the $h$-step ahead prediction for the GE-NoVaS-without-$a_0$ method}
\label{algori1}
\centering
  \centering
  \begin{tabular} {p{29pt}p{280pt}}   
    Step 1 & Define a grid of possible $\alpha$ values, $\{\alpha_k;~ k = 1,\cdots,K\}$. Fix $\alpha = \alpha_k$, then calculate the optimal combination of $\alpha_k,a_1,\cdots,a_p$ of the GE-NoVaS-without-$a_0$ method which minimizes $|Kurtosis(W_t)-3|$.\\
    Step 2 & Derive the analytic form of \cref{3e13} using $\alpha_k,a_1,\cdots,a_p$from the first step.\\
    Step 3 & Generate $\{W^*_{n+1},\cdots, W^*_{n+h}\}$ $M$ times from a standard normal distribution or the empirical distribution $\hat{F}_w$. Plug $\{W^*_{n+1},\cdots, W^*_{n+h}\}$ into the analytic form of \cref{3e13} to obtain $M$ pseudo-values $\{\hat{Y}^{*}_{n+h,1},\cdots,\hat{Y}^{*}_{n+h,M}\}$.\\
    Step 4 & Calculate the optimal predictor of $g(Y_{n+h})$ by taking the sample mean (under $L_2$ risk criterion) or sample median (under $L_1$ risk criterion) of the set $\{g(\hat{Y}^{*}_{n+h,1}),\cdots,g(\hat{Y}^{*}_{n+h,M})\}$.\\
    Step 5 & Repeat above steps with different $\alpha$ values from $\{\alpha_k;~ k = 1,\cdots,K\}$ to get $K$ prediction results.   \\ 
  \end{tabular}
\end{algorithm}

\subsection{The potential instability of the GE-NoVaS method}\label{susec:potentialinstability}
\noindent 
Next, we provide an illustration to compare GE-NoVaS and GE-NoVaS-without-$a_0$ methods on predicting volatility of Microsoft Corporation (MSFT) daily closing price from January 8, 1998 to December 31,1999 and show an interesting finding that the long-term time-aggregated predictions of GE-NoVaS method is unstable under $L_2$ criterion. Based on the finding of \citet{awartani2005predicting}, squared log-returns can be used as a proxy for volatility to render a correct ranking of different GARCH models in terms of a quadratic loss function. Log-returns series $\{Y_t\}$ can be computed by the equation shown below:
\begin{equation}
    Y_t = 100\times log(X_{t+1}/X_t)  \label{Eq.213}
\end{equation}
where, $\{X_{t}\}$ is the corresponding MSFT daily closing price series. For achieving a comprehensive comparison, we use 250 financial log-returns as a sliding-window to do POOS 1-step, 5-steps and 30-steps (long-term) ahead time-aggregated predictions under $L_2$ criterion. Then, we roll this window through whole dataset, i.e., we use $\{Y_1,\cdots,Y_{250}\}$ to predict $Y_{251}^2,\{Y_{251}^2,\cdots,Y_{255}^2\}$ and $\{Y_{251}^2,\cdots,Y_{280}^2\}$; then use $\{Y_2,\cdots,Y_{251}\}$ to predict $Y_{252}^2,\{Y_{252}^2,\cdots,Y_{256}^2\}$ and $\{Y_{252}^2,\cdots,Y_{281}^2\}$, for 1-step, 5-steps and 30-steps aggregated predictions respectively, and so on. We can define all 1-step, 5-steps and 30-steps ahead time-aggregated predictions as $\{\hat{Y}_{k,1}^2\}$, $\{\hat{Y}_{i,5}^2\}$ and $\{\hat{Y}_{j,30}^2\}$ which are presented as below:
\begin{equation}
\begin{split}
&\text{Assume there are total}~N~ \text{log-returns data points:}\\
    &\hat{Y}_{k,1}^2 = \hat{Y}_{k+1}^2,~k=250,251,\cdots, N - 1\\
    &\hat{Y}_{i,5}^2 = \sum_{m=1}^5\hat{Y}^2_{i+m},~i = 250,251,\cdots, N - 5\\
    &\hat{Y}_{j,30}^2 =\sum_{m=1}^{30}\hat{Y}^2_{j+m},~j = 250,251,\cdots, N - 30\\
 \label{4e17}
\end{split}
\end{equation}

In \cref{4e17}, $\hat{Y}_{k+1}^2,\hat{Y}_{i+m}^2,\hat{Y}_{j+m}^2$ are single-step predictions of squared log-returns by two NoVaS-type methods. To get ``Prediction Errors'' for two methods, we can calculate the ``Loss'' by comparing aggregated prediction results with realized aggregated values based on the formula \cref{eq:4.1}:

\begin{equation}
    L_{p,h} = \sum_{p}(\hat{Y}_{p,h}^2-\sum_{m=1}^h(Y_{p+m}^2))^2,~~p \in \{k,i,j\};~~h \in \{1,5,30\} \label{eq:4.1}
\end{equation}
where, $\{Y_{p+m}^2\}$ are realized squared log-returns. To show the potential instability of the GE-NoVaS method under $L_{2}$ criterion, we take $\alpha$ to be 0.5 to build a toy example. In the algorithm of performing the GE-NoVaS method, $\alpha$ could take an optimal value from a discrete set $\{0.1,\cdots,0.8\}$ based on prediction performance. 

\begin{figure}[htbp]
\centering
\includegraphics[width=14cm,height=7cm]{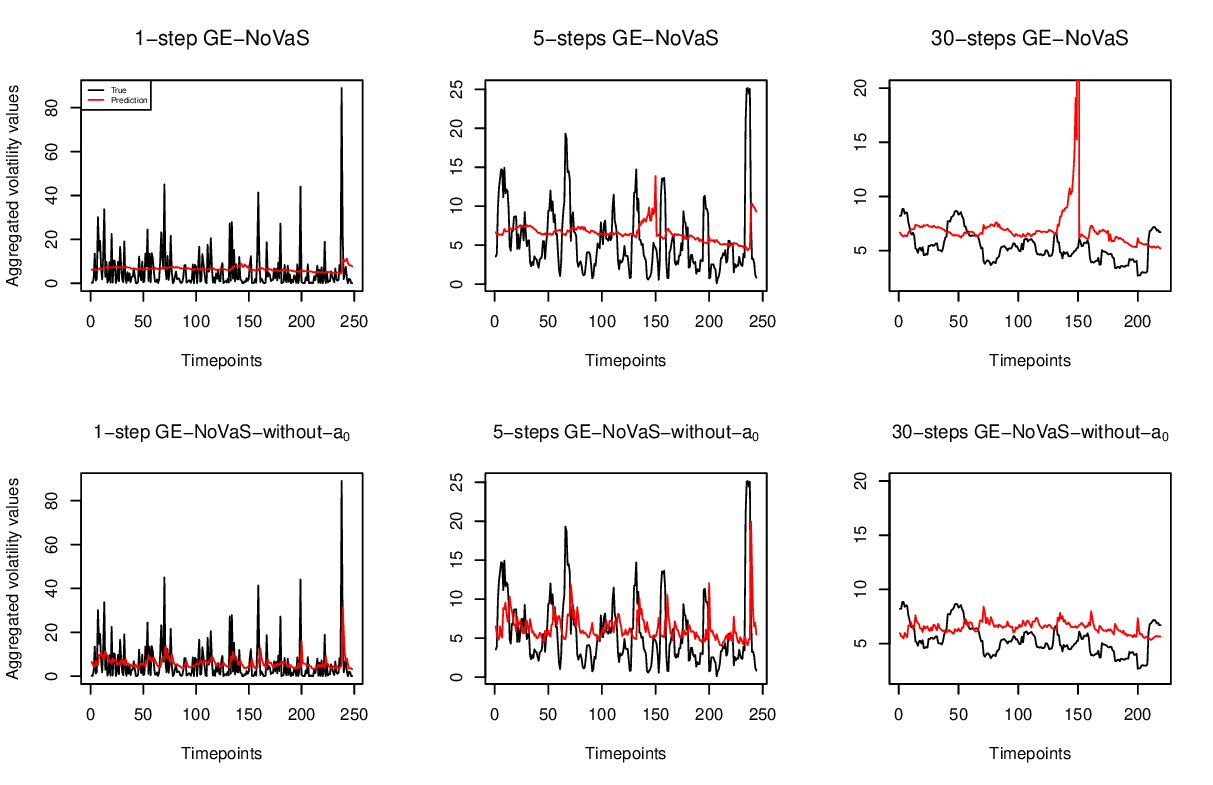}
\caption{Curves of the true and predicted time-aggregated squared log-returns from GE-NoVaS and GE-NoVaS-without-$a_0$ methods}
\label{2f1}
\end{figure}

From \cref{2f1}, we can clearly find that the GE-NoVaS-without-$a_0$ method can better capture true features. On the other hand, the GE-NoVaS method returns unstable results for 30-steps ahead time-aggregated predictions.

\section{Data analysis and results}\label{sec:analysisandresults}
\noindent To perform extensive data analysis in a bid to validate our method, we deploy POOS predictions using two NoVaS and standard GARCH(1,1) methods with simulated and real-world data. All results are collated in \cref{3t1}. For controlling the dependence of prediction performances on the length of the dataset, we build datasets with two fixed lengths--250 or 500--to mimic 1-year or 2-years data, respectively. At the same time, we choose the window-size for our rollover forecasting analysis to be 100 or 250 for 1-year or 2-years datasets.
\subsection{Simulation study}
\noindent We use same simulation Models 1-4 from \cite{chen2019optimal} shown as below to mimic four 1-year datasets. Recall that one NoVaS method can generate $L_1$ or $L_2$ predictor and $\{W^{*}\}$ can be chosen from a normal distribution or empirical distribution, thus there are four variants of one specific NoVaS method. We take the best-performing result among four variants of a specific NoVaS method to be its final prediction. Finally, we keep applying the formula \cref{eq:4.1} to measure the performance of different methods like we did in \cref{susec:potentialinstability}. \\
\\
\textbf{Model 1:} Time-varying GARCH(1,1) with Gaussian errors\\
$X_t = \sigma_t\epsilon_t,~\sigma_t^2 = 0.00001 + \beta_{1,t}\sigma_{t-1}^2+\alpha_{1,t}X_{t-1}^2,~\{\epsilon_t\}\sim i.i.d.~N(0,1)$\\
$\alpha_{1,t} = 0.1 - 0.05t/n$; $\beta_{1,t} = 0.73 + 0.2t/n,~n = 250$\\
\textbf{Model 2:} Standard GARCH(1,1) with Gaussian errors\\
$X_t = \sigma_t\epsilon_t,~\sigma_t^2 = 0.00001 + 0.73\sigma_{t-1}^2+0.1X_{t-1}^2,~\{\epsilon_t\}\sim i.i.d.~N(0,1)$\\
\textbf{Model 3:} (Another) Standard GARCH(1,1) with Gaussian errors\\
$X_t = \sigma_t\epsilon_t,~\sigma_t^2 = 0.00001 + 0.8895\sigma_{t-1}^2+0.1X_{t-1}^2,~\{\epsilon_t\}\sim i.i.d.~N(0,1)$\\
\textbf{Model 4:} Standard GARCH(1,1) with Student-$t$ errors\\
$X_t = \sigma_t\epsilon_t,$ $~\sigma_t^2 = 0.00001 + 0.73\sigma_{t-1}^2+0.1X_{t-1}^2,$\\ $~\{\epsilon_t\}\sim i.i.d.~t$ $\text{distribution with five degrees of freedom}$\\
\\
\textbf{Result analysis:} From the first block of \cref{3t1}, we can read that both NoVaS methods are superior than GARCH(1,1) model. Even these simulated datasets are generated from GARCH(1,1)-type models, GE-NoVaS-without-$a_0$ method can bring around 66$\%$ and 48$\%$ improvements compared to  GARCH(1,1) model on 5-steps ahead time-aggregated predictions of Model-4 and Model-1 data, respectively. Notably, GARCH(1,1) brings terrible results for 30-steps ahead time-aggregated predictions of Model-4 simulated data. Take a closer look at these results, we can observe that almost all optimal results come from applying the GE-NoVaS-without-$a_0$ method. Moreover, GE-NoVaS method is beaten by GARCH(1,1) on forecasting 30-steps ahead time-aggregated Model-2 data. On the other hand, the GE-NoVaS-without-$a_0$ method keep consistently stable results. Additionally, with simulation implementations, the ability against model misspecification of NoVaS methods is checked in \ref{apdix:A}.

\subsection{A few real datasets}
\noindent We also present a variety of real-world datasets of different size and intrinsic behavior
\begin{itemize}
    \item 2 year period data: 2018$\sim$2019 Stock price data.
    \item 1 year period data: 2019 Stock price and Index data.
    \item 1 year period volatile data due to pandemic: 11.2019$\sim$10.2020 Stock price, Currency and Index data.
\end{itemize}
Taking into account three types of real-world data is made to challenge our new method and explore the existing method in different regimes. We also tactically pay more attention to short and volatile data since it is a harder task to handle. \cref{Eq.213} is continually used to get log-returns series of different datasets.  
\\
\\
\textbf{Result analysis:} From last three blocks of \cref{3t1}, there is no optimal result which comes from the GARCH(1,1) method. When target data is short and volatile, GARCH(1,1) gives terrible results for 30-steps ahead time-aggregated predictions, such as volatile Djones, CADJPY and IBM cases. Within two NoVaS methods, GE-NoVaS-without-$a_0$ method outperforms GE-NoVaS method for three types of real-world data. More specifically, around 70$\%$ and 30$\%$ improvements are created by our new method compared to the existing GE-NoVaS method on forecasting 30-steps ahead time-aggregated volatile Djones and CADJPY data, respectively. We should also notice that the GE-NoVaS method is again beaten by GARCH(1,1) model on 30-steps ahead aggregated predictions of 2018$\sim$2019 BAC data. On the other hand, GE-NoVaS-without-$a_0$ method stands stably. See \ref{apdix:A} for more results.
\begin{table}[htbp]
    \begin{adjustbox}{width=1\textwidth}
\footnotesize
\begin{tabular}{llcccc}
  \toprule 
 & & \thead{\footnotesize GE-NoVaS} & \thead{ \footnotesize GE-NoVaS-without-$a_0$} & \thead{ \footnotesize GARCH(1,1)}  & \thead{ \footnotesize P-value(CW-test)} \\
  \midrule
 \multirow{12}{*}{\rotatebox[origin=c]{90}{Simulated-1-year-data}} & Model-1-1step & 0.91369 & \textbf{0.88781}  & 1.00000  &\\
   & Model-1-5steps & 0.61001 & \textbf{0.52872} & 1.00000 &\\
   & Model-1-30steps & 0.77250 & \textbf{0.73604} & 1.00000 &\\
   & Model-2-1step & 0.97796 & \textbf{0.94635} & 1.00000 &\\
   & Model-2-5steps & 0.98127 & \textbf{0.96361}  & 1.00000 &\\
   & Model-2-30steps & 1.38353 & \textbf{0.98872} & 1.00000 &\\
   & Model-3-1step & 0.99183 & \textbf{0.92829}  & 1.00000 &\\
   & Model-3-5steps & 0.77088 & \textbf{0.67482}  & 1.00000 &\\
   & Model-3-30steps & 0.79672 & \textbf{0.71003}  & 1.00000 &\\
   & Model-4-1step & 0.83631 & \textbf{0.78087}  & 1.00000 &\\
   & Model-4-5steps & 0.38296 & \textbf{0.34396}  & 1.00000 &\\
   & Model-4-30steps & \textbf{0.00199} & 0.00201  & 1.00000 &\\\hline
  \multirow{6}{*}{\rotatebox[origin=c]{90}{2-years-data}}  & 2018$\sim$2019-MCD-1step & 0.99631  & \textbf{0.99614}  & 1.00000 & 0.00053\\
   & 2018$\sim$2019-MCD-5steps & 0.95403 & \textbf{0.92120} &  1.00000 &0.03386\\
   & 2018$\sim$2019-MCD-30steps & 0.75730 & \textbf{0.62618} & 1.00000 & 0.19691\\
   & 2018$\sim$2019-BAC-1step & 0.98393 & \textbf{0.97966}  & 1.00000 & 0.09568\\
   & 2018$\sim$2019-BAC-5steps & 0.98885 & \textbf{0.95124}  & 1.00000 &0.07437\\
   & 2018$\sim$2019-BAC-30steps & 1.14111 & \textbf{0.87414}  & 1.00000 & 0.03643\\ \hline
   \multirow{9}{*}{\rotatebox[origin=c]{90}{1-year-data}}& 2019-AAPL-1step & 0.84533 & \textbf{0.80948}   & 1.00000 & 0.25096\\
   & 2019-AAPL-5steps & 0.85401 & \textbf{0.68191}  & 1.00000 & 0.06387\\
   & 2019-AAPL-30steps & 0.99043 & \textbf{0.73823}  & 1.00000 & 0.17726\\
   & 2019-Djones-1step & 0.96752 & \textbf{0.96365}   & 1.00000 & 0.34514\\
   & 2019-Djones-5steps & 0.98725 & \textbf{0.89542}  & 1.00000 & 0.24529\\
   & 2019-Djones-30steps & 0.86333 & \textbf{0.80304}  & 1.00000 & 0.23766\\
   & 2019-SP500-1step & 0.96978 & \textbf{0.92183}  & 1.00000 & 0.45693\\
   & 2019-SP500-5steps & 0.96704 & \textbf{0.75579}  & 1.00000 & 0.24402\\
   & 2019-SP500-30steps & 0.34389 & \textbf{0.29796}  & 1.00000 &0.08148 \\ \hline
   \multirow{12}{*}{\rotatebox[origin=c]{90}{Volatile-1-year-data}}& 11.2019$\sim$10.2020-IBM-1step & \textbf{0.80222} & 0.80744  &  1.00000 & 0.16568\\
   & 11.2019$\sim$10.2020-IBM-5steps & \textbf{0.38933} & 0.40743  &  1.00000 & 0.03664\\
   & 11.2019$\sim$10.2020-IBM-30steps & 0.01143 & \textbf{0.00918}   & 1.00000 &0.15364\\
   & 11.2019$\sim$10.2020-CADJPY-1step & \textbf{0.46940} & 0.48712  & 1.00000 & 0.16230\\
   & 11.2019$\sim$10.2020-CADJPY-5steps & \textbf{0.11678} & 0.13549  & 1.00000 & 0.06828\\
   & 11.2019$\sim$10.2020-CADJPY-30steps & 0.00584 & \textbf{0.00394} & 1.00000 & 0.15174\\
   & 11.2019$\sim$10.2020-SP500-1step & 0.97294 & \textbf{0.92349}  & 1.00000 & 0.05536\\
   & 11.2019$\sim$10.2020-SP500-5steps & 0.96590 & \textbf{0.75183} & 1.00000 & 0.17380\\
   & 11.2019$\sim$10.2020-SP500-30steps & 0.34357 & \textbf{0.29793}  & 1.00000 &0.16022\\
   & 11.2019$\sim$10.2020-Djones-1step & \textbf{0.56357} & 0.57550  & 1.00000 &0.11099\\
   & 11.2019$\sim$10.2020-Djones-5steps & \textbf{0.09810} & 0.11554 & 1.00000 & 0.45057\\
   & 11.2019$\sim$10.2020-Djones-30steps & 4.32E-05 & \textbf{1.24E-05} & 1.00000 &0.68487\\
       \bottomrule
    \end{tabular}
    \end{adjustbox}\\
    \footnotesize
     \textit{Note:} The values presented in GE-NoVaS and GE-NoVaS-without-$a_0$ columns are relative performance compared with `standard' GARCH(1,1) method. The null hypothesis of the CW-test is that parsimonious and larger models have equal mean squared prediction error (MSPE). The alternative is that the larger model has a smaller MSPE. 
          \caption{Comparisons of different methods' forecasting performance}
            \label{3t1}
\end{table}

\subsection{Statistical significance}
\noindent However, one may think the victory of our new methods is just specific to these samples. Therefore, we challenge this superiority by testing statistical significance. Noticing the GE-NoVaS-without-$a_0$ method is the nested method (taking $a_0=0$ in the larger model) compared with the GE-NoVaS method, we deploy the CW-test  \citep{clark2007approximately} to make sure that the removing-$a_0$ idea is also statistically reasonable, see the $P$-value column in the \cref{3t1} for tests' results\footnote{The reason of not performing CW-tests on simulation cases is that each prediction performance of simulation is the average value of 5 replications.}. These CW-tests results imply that the null-hypothesis should not be rejected for almost all cases under 5$\%$ level of significance, which advocates equivalence of new method to the existing one.

\subsection{Summary}
\noindent We summarize our findings as follow:
\begin{itemize}
    \item Existing GE-NoVaS and new GE-NoVaS-without-$a_0$ methods provide substantial improvement for time-aggregated prediction which hints to stability of NoVaS-type methods for providing long-horizon inferences. 
    \item Our new method has a superior performance than the GE-NoVaS method, especially for shorter sample size or more volatile data. This is significant given GARCH-type models are difficult to estimate in shorter samples. 
    \item We provide a statistical hypothesis test that votes for our model advocating a more parsimonious fit, especially for long-term time-aggregated predictions. 
\end{itemize}

\section{Discussion }\label{conclusion}
\noindent In this article, we explored the GE-NoVaS method toward short and long time-aggregated predictions and proposed a new variant that is based on a parsimonious model, has a better empirical performance and yet is statistically reasonable. We hope these empirical findings open up avenues where one can explore other specific transformation structures to improve existing forecasting frameworks.

\bibliography{aggfore}

\begin{thebibliography}{14}
\expandafter\ifx\csname natexlab\endcsname\relax\def\natexlab#1{#1}\fi
\providecommand{\url}[1]{\texttt{#1}}
\providecommand{\href}[2]{#2}
\providecommand{\path}[1]{#1}
\providecommand{\DOIprefix}{doi:}
\providecommand{\ArXivprefix}{arXiv:}
\providecommand{\URLprefix}{URL: }
\providecommand{\Pubmedprefix}{pmid:}
\providecommand{\doi}[1]{\href{http://dx.doi.org/#1}{\path{#1}}}
\providecommand{\Pubmed}[1]{\href{pmid:#1}{\path{#1}}}
\providecommand{\bibinfo}[2]{#2}
\ifx\xfnm\relax \def\xfnm[#1]{\unskip,\space#1}\fi
%Type = Article
\bibitem[{Awartani and Corradi(2005)}]{awartani2005predicting}
\bibinfo{author}{Awartani, B.M.}, \bibinfo{author}{Corradi, V.},
  \bibinfo{year}{2005}.
\newblock \bibinfo{title}{Predicting the volatility of the s\&p-500 stock index
  via garch models: the role of asymmetries}.
\newblock \bibinfo{journal}{International Journal of forecasting}
  \bibinfo{volume}{21}, \bibinfo{pages}{167--183}.
%Type = Article
\bibitem[{Bansal et~al.(2016)Bansal, Kiku and Yaron}]{bansal2016risks}
\bibinfo{author}{Bansal, R.}, \bibinfo{author}{Kiku, D.},
  \bibinfo{author}{Yaron, A.}, \bibinfo{year}{2016}.
\newblock \bibinfo{title}{Risks for the long run: Estimation with time
  aggregation}.
\newblock \bibinfo{journal}{Journal of Monetary Economics}
  \bibinfo{volume}{82}, \bibinfo{pages}{52--69}.
%Type = Phdthesis
\bibitem[{Chen(2018)}]{chen2018prediction}
\bibinfo{author}{Chen, J.}, \bibinfo{year}{2018}.
\newblock \bibinfo{title}{Prediction in Time Series Models and Model-free
  Inference with a Specialization in Financial Return Data}.
\newblock Ph.D. thesis. UC San Diego.
%Type = Article
\bibitem[{Chen and Politis(2019)}]{chen2019optimal}
\bibinfo{author}{Chen, J.}, \bibinfo{author}{Politis, D.N.},
  \bibinfo{year}{2019}.
\newblock \bibinfo{title}{Optimal multi-step-ahead prediction of arch/garch
  models and novas transformation}.
\newblock \bibinfo{journal}{Econometrics} \bibinfo{volume}{7},
  \bibinfo{pages}{34}.
%Type = Article
\bibitem[{Chud{\`y} et~al.(2020)Chud{\`y}, Karmakar and Wu}]{chudy2020long}
\bibinfo{author}{Chud{\`y}, M.}, \bibinfo{author}{Karmakar, S.},
  \bibinfo{author}{Wu, W.B.}, \bibinfo{year}{2020}.
\newblock \bibinfo{title}{Long-term prediction intervals of economic time
  series}.
\newblock \bibinfo{journal}{Empirical Economics} \bibinfo{volume}{58},
  \bibinfo{pages}{191--222}.
%Type = Article
\bibitem[{Clark and West(2007)}]{clark2007approximately}
\bibinfo{author}{Clark, T.E.}, \bibinfo{author}{West, K.D.},
  \bibinfo{year}{2007}.
\newblock \bibinfo{title}{Approximately normal tests for equal predictive
  accuracy in nested models}.
\newblock \bibinfo{journal}{Journal of econometrics} \bibinfo{volume}{138},
  \bibinfo{pages}{291--311}.
%Type = Article
\bibitem[{Engle(1982)}]{engle1982autoregressive}
\bibinfo{author}{Engle, R.F.}, \bibinfo{year}{1982}.
\newblock \bibinfo{title}{Autoregressive conditional heteroscedasticity with
  estimates of the variance of united kingdom inflation}.
\newblock \bibinfo{journal}{Econometrica: Journal of the Econometric Society} ,
  \bibinfo{pages}{987--1007}.
%Type = Article
\bibitem[{Fryzlewicz et~al.(2008)Fryzlewicz, Sapatinas, Rao
  et~al.}]{fryzlewicz2008normalized}
\bibinfo{author}{Fryzlewicz, P.}, \bibinfo{author}{Sapatinas, T.},
  \bibinfo{author}{Rao, S.S.}, et~al., \bibinfo{year}{2008}.
\newblock \bibinfo{title}{Normalized least-squares estimation in time-varying
  arch models}.
\newblock \bibinfo{journal}{The Annals of Statistics} \bibinfo{volume}{36},
  \bibinfo{pages}{742--786}.
%Type = Article
\bibitem[{Gulay and Emec(2018)}]{gulay2018comparison}
\bibinfo{author}{Gulay, E.}, \bibinfo{author}{Emec, H.}, \bibinfo{year}{2018}.
\newblock \bibinfo{title}{Comparison of forecasting performances: Does
  normalization and variance stabilization method beat garch (1, 1)-type
  models? empirical evidence from the stock markets}.
\newblock \bibinfo{journal}{Journal of Forecasting} \bibinfo{volume}{37},
  \bibinfo{pages}{133--150}.
%Type = Article
\bibitem[{Karmakar et~al.(2020)Karmakar, Chudy and Wu}]{karmakar2020long}
\bibinfo{author}{Karmakar, S.}, \bibinfo{author}{Chudy, M.},
  \bibinfo{author}{Wu, W.B.}, \bibinfo{year}{2020}.
\newblock \bibinfo{title}{Long-term prediction intervals with many covariates}.
\newblock \bibinfo{journal}{To appear at Journal of Time-series Analysis. arXiv
  preprint arXiv:2012.08223} .
%Type = Article
\bibitem[{Kitsul and Wright(2013)}]{kitsul2013economics}
\bibinfo{author}{Kitsul, Y.}, \bibinfo{author}{Wright, J.H.},
  \bibinfo{year}{2013}.
\newblock \bibinfo{title}{The economics of options-implied inflation
  probability density functions}.
\newblock \bibinfo{journal}{Journal of Financial Economics}
  \bibinfo{volume}{110}, \bibinfo{pages}{696--711}.
%Type = Incollection
\bibitem[{Politis(2003)}]{politis2003normalizing}
\bibinfo{author}{Politis, D.N.}, \bibinfo{year}{2003}.
\newblock \bibinfo{title}{A normalizing and variance-stabilizing transformation
  for financial time series}, \bibinfo{publisher}{Elsevier Inc.}
%Type = Incollection
\bibitem[{Politis(2015)}]{politis2015modelfreepredictionprinciple}
\bibinfo{author}{Politis, D.N.}, \bibinfo{year}{2015}.
\newblock \bibinfo{title}{The model-free prediction principle}, in:
  \bibinfo{booktitle}{Model-Free Prediction and Regression}.
  \bibinfo{publisher}{Springer}, pp. \bibinfo{pages}{13--30}.
%Type = Article
\bibitem[{Starica(2003)}]{starica2003garch}
\bibinfo{author}{Starica, C.}, \bibinfo{year}{2003}.
\newblock \bibinfo{title}{Is garch (1, 1) as good a model as the accolades of
  the nobel prize would imply?}
\newblock \bibinfo{journal}{Available at SSRN 637322} .

\end{thebibliography}

\clearpage

\appendix
\section{Additional simulation study and data analysis results}\label{apdix:A}
\subsection{Additional simulation study: model misspecification}
\noindent In the real world, it is hard to convincingly say if the data obeys one particular type of GARCH model, so we want to provide four more GARCH-type models to simulate one-year datasets to see if our methods are satisfactory no matter what the underlying distribution and GARCH-type model are. Simulation study results are presented in \cref{Table:appendixtAable1}, which imply NoVaS-type methods are more robust against model misspecification and the GE-NoVaS-without-$a_0$ is the best method.
\\
\\
\textbf{Model 5:} Another Time-varying GARCH(1,1) with Gaussian errors\\
$X_t = \sigma_t\epsilon_t,~\sigma_t^2 = \omega_{0,t} + \beta_{1,t}\sigma_{t-1}^2+\alpha_{1,t}X_{t-1}^2,~\{\epsilon_t\}\sim i.i.d.~N(0,1)$\\
$g_t = t/n; \omega_{0,t}= -4sin(0.5\pi g_t)+5; \alpha_{1,t} = -1(g_t-0.3)^2 + 0.5; \beta_{1,t} = 0.2sin(0.5\pi g_t)+0.2,~n = 250$\\
\textbf{Model 6:} Exponential GARCH(1,1) with Gaussian errors\\
$X_t = \sigma_t\epsilon_t,~\log(\sigma_t^2) = 0.00001 + 0.8895\log(\sigma^2_{t-1})+0.1\epsilon_{t-1}+0.3(\abs{\epsilon_{t-1}}-E\abs{\epsilon_{t-1}}),$\\$~\{\epsilon_t\}\sim i.i.d.~N(0,1)$\\
\textbf{Model 7:} GJR-GARCH(1,1) with Gaussian errors\\
$X_t = \sigma_t\epsilon_t,~\sigma_t^2 = 0.00001 + 0.5\sigma^2_{t-1}+0.5X_{t-1}^2-0.5I_{t-1}X_{t-1}^2,~\{\epsilon_t\}\sim i.i.d.~N(0,1)\\
I_{t} = 1~\text{if}~ X_t \leq 0; I_{t} = 0~ \text{otherwise}$\\
\textbf{Model 8:} Another GJR-GARCH(1,1) with Gaussian errors\\
$X_t = \sigma_t\epsilon_t,~\sigma_t^2 = 0.00001 + 0.73\sigma^2_{t-1}+0.1X_{t-1}^2+0.3I_{t-1}X_{t-1}^2,~\{\epsilon_t\}\sim i.i.d.~N(0,1)\\
I_{t} = 1~\text{if}~ X_t \leq 0; I_{t} = 0~ \text{otherwise}$\\

\begin{longtable}{llccc}
\toprule
& & \thead{GE-NoVaS} & \thead{ GE-NoVaS-without-$a_0$} & \thead{ GARCH(1,1)} \\
\hline
 &   M5-1step & 0.91538 & \textbf{0.83168} & 1.00000 \\
 &   M5-5steps & 0.49169 & \textbf{0.43772} &  1.00000 \\
 &   M5-30steps & 0.25009 & \textbf{0.22659} &  1.00000 \\
 &   M6-1step & 0.95939 & \textbf{0.94661} & 1.00000 \\
 &   M6-5steps & 0.93594 & \textbf{0.84719} & 1.00000 \\
 &   M6-30steps & 0.84401 & \textbf{0.70301} &  1.00000 \\
 &   M7-1step & 0.84813 & \textbf{0.73553} &  1.00000 \\
 &   M7-5steps & 0.50849 & \textbf{0.46618} & 1.00000 \\
 &   M7-30steps & 0.06832 & \textbf{0.06479} &  1.00000 \\
 &   M8-1step & 0.79561 & \textbf{0.76586} &  1.00000 \\
 &   M8-5steps & 0.48028 & \textbf{0.38107} & 1.00000 \\
 &   M8-30steps & 0.00977 & \textbf{0.00918} &  1.00000 \\
    \hline
          \caption{Comparisons of different methods' forecasting performance on simulated-1-year-data}
\label{Table:appendixtAable1}
\end{longtable}

\subsection{Additional data analysis: 1-year datasets}
\noindent For making our data analysis more comprehensive, we present more results of predictions on 1-year real-world datasets in \cref{Table:appendixtAable2}. One interesting finding is that the GE-NoVaS method is significantly beaten by using GARCH(1,1) model for some cases, such as BAC, TSLA and Smallcap datasets. The GE-NoVaS-without-$a_0$ method still keeps great forecasting performance. 

\begin{longtable}{llccc}
\toprule
& & \thead{GE-NoVaS} & \thead{ GE-NoVaS-without-$a_0$} & \thead{ GARCH(1,1)} \\
\hline
 &   2019-MCD-1step & 0.95959 & \textbf{0.93141} &  1.00000 \\
 &   2019-MCD-5steps & 1.00723 & \textbf{0.90061} &  1.00000 \\
 &   2019-MCD-30steps & 1.05239 & \textbf{0.80805} &  1.00000 \\
 &   2019-BAC-1step & 1.04272 & \textbf{0.97757} & 1.00000 \\
 &   2019-BAC-5steps & 1.22761 & \textbf{0.89571} &  1.00000 \\
 &   2019-BAC-30steps & 1.45020 & 1.01175 &  \textbf{1.00000} \\
 &   2019-MSFT-1step & 1.03308 & \textbf{0.98469} &  1.00000 \\
 &   2019-MSFT-5steps & 1.22340 & 1.02387 &  \textbf{1.00000} \\
 &   2019-MSFT-30steps & 1.23020 & \textbf{0.97585} & 1.00000 \\
 &   2019-TSLA-1step & 1.00428 & \textbf{0.98646} & 1.00000 \\
 &   2019-TSLA-5steps & 1.06610 & \textbf{0.97523} &  1.00000 \\
 &   2019-TSLA-30steps & 2.00623 & \textbf{0.87158} & 1.00000 \\
 &   2019-Bitcoin-1step & 0.89929 & \textbf{0.86795} &  1.00000 \\
 &   2019-Bitcoin-5steps & 0.62312 & \textbf{0.55620} & 1.00000 \\
 &   2019-Bitcoin-30steps & 0.00733 & \textbf{0.00624} &  1.00000 \\
 &   2019-Nasdaq-1step & 0.99960 & \textbf{0.93558} &  1.00000 \\
 &   2019-Nasdaq-5steps & 1.15282 & \textbf{0.84459} &  1.00000 \\
 &   2019-Nasdaq-30steps & 0.68994 & \textbf{0.58924} &  1.00000 \\
 &   2019-NYSE-1step & 0.92486 & \textbf{0.90407} & 1.00000 \\
 &   2019-NYSE-5steps & 0.86249 & \textbf{0.69822} & 1.00000 \\
 &   2019-NYSE-30steps & 0.22122 & \textbf{0.18173} &  1.00000 \\
 &   2019-Smallcap-1step & 1.02041 & \textbf{0.98731} & 1.00000 \\
 &   2019-Smallcap-5steps & 1.15868 & \textbf{0.87700} &  1.00000 \\
 &   2019-Samllcap-30steps & 1.30467 & \textbf{0.88825} & 1.00000 \\
 &   2019-BSE-1step & 0.70667 & \textbf{0.67694} &  1.00000 \\
 &   2019-BSE-5steps & 0.25675 & \textbf{0.23665} & 1.00000 \\
 &   2019-BSE-30steps & 0.03764 & \textbf{0.02890} &  1.00000 \\
 &   2019-BIST-1step & 0.96807 & \textbf{0.95467} & 1.00000 \\
 &   2019-BIST-5steps & 0.98944 & \textbf{0.82898} &1.00000 \\
 &   2019-BIST-30steps & 2.21996 &\textbf{ 0.88511} & 1.00000 \\
    \hline
          \caption{Comparisons of different methods' forecasting performance on real-world-1-year-data}
\label{Table:appendixtAable2}
\end{longtable}

\subsection{Additional data analysis: volatile 1-year datasets}
\noindent Similarly, we consider more volatile 1-year datasets. All prediction results are tabulated in \cref{Table:appendixtBable3}. It is clear that both NoVaS-type methods still outperform the GARCH(1,1) model for short- and long-term time-aggregated forecasting. Although the GE-NoVaS method attaches optimal performance in some cases, we should notice that the GE-NoVaS-without-$a_0$ method still gives almost same but slightly worse results. Interestingly, the GE-NoVaS-without-$a_0$ method can introduce significant improvement compared with the GE-NoVaS method for 30-steps ahead predictions. This again hints to more robustness of our new method specifically for long-term aggregated predictions.  
\begin{longtable}{llccc}
\toprule
& & \thead{GE-NoVaS} & \thead{ GE-NoVaS-without-$a_0$} & \thead{ GARCH(1,1)} \\
\hline
 &    11.2019$\sim$10.2020-MCD-1step & \textbf{0.51755} & 0.58018 &  1.00000 \\
 &    11.2019$\sim$10.2020-MCD-5steps & \textbf{0.10725} & 0.17887 &  1.00000 \\
 &    11.2019$\sim$10.2020-MCD-30steps & 3.32E-05 & \textbf{7.48E-06} &  1.00000 \\
 &    11.2019$\sim$10.2020-AMZN-1step & 0.97099 & \textbf{0.90200} & 1.00000 \\
 &    11.2019$\sim$10.2020-AMZN-5steps & 0.88705 & \textbf{0.71789} &  1.00000\\
 &    11.2019$\sim$10.2020-AMZN-30steps & 0.58124& \textbf{0.53460} & 1.00000\\
 &    11.2019$\sim$10.2020-SBUX-1step & \textbf{0.68206} & 0.69943 & 1.00000 \\
 &    11.2019$\sim$10.2020-SBUX-5steps & \textbf{0.24255} & 0.30528 & 1.00000 \\
 &    11.2019$\sim$10.2020-SBUX-30steps & 0.00499 & \textbf{0.00289} &1.00000 \\
 &    11.2019$\sim$10.2020-MSFT-1step & \textbf{0.80133} & 0.84502 &  1.00000 \\
 &    11.2019$\sim$10.2020-MSFT-5steps & \textbf{0.35567} & 0.37528 & 1.00000 \\
 &    11.2019$\sim$10.2020-MSFT-30steps & 0.01342 & \textbf{0.00732} & 1.00000 \\
 &    11.2019$\sim$10.2020-EURJPY-1step & 0.95093 & \textbf{0.94206} & 1.00000 \\
 &    11.2019$\sim$10.2020-EURJPY-5steps & \textbf{0.76182} & 0.76727 & 1.00000 \\
 &    11.2019$\sim$10.2020-EURJPY-30steps & 0.16202 & \textbf{0.15350} & 1.00000 \\
 &    11.2019$\sim$10.2020-CNYJPY-1step & \textbf{0.77812} & 0.79877 & 1.00000 \\
 &    11.2019$\sim$10.2020-CNYJPY-5steps & \textbf{0.38875} & 0.40569 & 1.00000 \\
 &    11.2019$\sim$10.2020-CNYJPY-30steps & 0.08398 & \textbf{0.06270} & 1.00000 \\
 &    11.2019$\sim$10.2020-Smallcap-1step & \textbf{0.58170} & 0.60931 &  1.00000 \\
 &    11.2019$\sim$10.2020-Smallcap-5steps & \textbf{0.10270} & 0.10337 & 1.00000 \\
 &    11.2019$\sim$10.2020-Smallcap-30steps & 7.00E-05 & \textbf{5.96E-05} &  1.00000 \\
 &    11.2019$\sim$10.2020-BSE-1step & \textbf{0.39493} & 0.39745 &  1.00000 \\
 &    11.2019$\sim$10.2020-BSE-5steps & \textbf{0.03320} & 0.04109 &  1.00000 \\
 &    11.2019$\sim$10.2020-BSE-30steps & 2.45E-05 & \textbf{1.82E-05} &  1.00000 \\
 &    11.2019$\sim$10.2020-NYSE-1step & \textbf{0.55741} & 0.57174 & 1.00000 \\
 &    11.2019$\sim$10.2020-NYSE-5steps & \textbf{0.08994} & 0.10182 &  1.00000 \\
 &    11.2019$\sim$10.2020-NYSE-30steps & 1.36E-05 & \textbf{6.64E-06} & 1.00000 \\
 &    11.2019$\sim$10.2020-USDXfuture-1step & 1.14621 & \textbf{0.99640} &  1.00000 \\
 &    11.2019$\sim$10.2020-USDXfuture-5steps & 0.61075 & \textbf{0.54834} & 1.00000 \\
 &    11.2019$\sim$10.2020-USDXfuture-30steps & 0.10723 & \textbf{0.10278} & 1.00000 \\
 &    11.2019$\sim$10.2020-Nasdaq-1step & \textbf{0.71380} & 0.75350 & 1.00000 \\
 &    11.2019$\sim$10.2020-Nasdaq-5steps & \textbf{0.29332} & 0.33519 & 1.00000 \\
 &    11.2019$\sim$10.2020-Nasdaq-30steps & 0.01223 & \textbf{0.00599} & 1.00000 \\
 &    11.2019$\sim$10.2020-Bovespa-1step & 0.60031 & \textbf{0.57558} &  1.00000 \\
 &    11.2019$\sim$10.2020-Bovespa-5steps & 0.08603 & \textbf{0.07447} & 1.00000 \\
 &    11.2019$\sim$10.2020-Bovespa-30steps & 6.87E-06 & \textbf{2.04E-06} & 1.00000 \\
    \hline
          \caption{Comparisons of different methods' forecasting performance on volatile-1-year-data}
\label{Table:appendixtBable3}
\end{longtable}

\end{document}